\documentclass[10pt]{wlscirep}
\usepackage{amssymb,amsmath,mathrsfs}
\usepackage{bm}
\usepackage{bm}
\usepackage{isomath}
\usepackage{siunitx}
\usepackage{ifthen}
\usepackage{soul}
\usepackage{cleveref}
\usepackage[nolist,nohyperlinks]{acronym}
\usepackage[normalem]{ulem}
\usepackage[colorinlistoftodos]{todonotes}
\newboolean{showrevision}
\setboolean{showrevision}{true}
\ifshowrevision %
	
	\newcommand{\delete}[1]{\textcolor{red}{\sout{#1}}}
\else
	
	\newcommand{\delete}[1]{}
\fi

\newcommand{\tightoverset}[2]{%
  \mathop{#2}\limits^{\vbox to -.5ex{\kern-0.75ex\hbox{\scriptsize{$#1$}}\vss}}}
\newcommand{\Gsum}{\tensor{\mathcal{G}}}
\newcommand{\Gneq}{\tensor{\mathcal{G}}^{\neq}}
\newcommand{\kparv}{{\bm{k}_{\parallel}}}
\newcommand{\kpar}{k_{\parallel}}
\newcommand{\sub}[1]{_{\mathrm{#1}}}
\newcommand{\unit}[1]{\,\,{\si{#1}}}

\newcommand{\alphat}{\tensor{\alpha}}
\newcommand{\unitvec}[1]{\hat{\bm{#1}}}

\newcommand{\tensor}[1]{{\tightoverset{\leftrightarrow}{#1}}}
\crefname{figure}{Fig.}{Figs.}
\crefname{equation}{equation}{equations}

\title{The local density of optical states of a metasurface}
\author[1]{Per Lunnemann}
\affil[1]{DTU Fotonik, Department of Photonics Engineering, \O stedsplads 343, DK-2800, Denmark}
\author[2,*]{A. Femius Koenderink}
\affil[2]{Center for Nanophotonics, FOM Institute AMOLF,
Science Park 104, 1098 XG Amsterdam, The Netherlands}
\affil[*]{fkoenderink@amolf.nl}
\begin{abstract}
While metamaterials  are often desirable for  near-field functions, such as perfect lensing, or cloaking, they are often quantified by their response to plane waves from the far field. Here, we present a theoretical analysis of the local density of states near lattices of discrete magnetic scatterers, i.e., the response to near field excitation by a point source.
Based on a point-dipole theory using Ewald summation and an array scanning method, we can swiftly and semi-analytically evaluate the local density of states (LDOS) for  magnetoelectric point sources in front of an infinite two-dimensional (2D) lattice composed of arbitrary magnetoelectric  dipole scatterers. The method takes into account radiation damping as well as all retarded electrodynamic interactions in a self-consistent manner.   We show that a lattice of magnetic scatterers evidences characteristic Drexhage oscillations. However, the oscillations are phase shifted relative to the electrically scattering lattice consistent with the difference expected for reflection off homogeneous magnetic respectively electric mirrors. Furthermore, we identify in which source-surface separation regimes the metasurface may be treated as a homogeneous interface, and in which homogenization fails. A strong frequency and in-plane position dependence of the LDOS close to the lattice reveals coupling to guided modes supported by the lattice.
\end{abstract}

\begin{document}
\flushbottom
\maketitle
\thispagestyle{empty}
\section*{Introduction}
%\com{No sections in Opt. Lett.}
Spontaneous emission is the irreversible decay of a quantum emitter upon emission of a photon that arises due to interaction with the fluctuating electromagnetic vacuum field. If the local density of available photon states (LDOS), and thereby of vacuum fluctuations, is modified the spontaneous emission rate may be inhibited or enhanced. This effect was discussed first for microcavities by Purcell in 1946 \cite{Purcell1946}. The effect was clearly demonstrated by Drexhage in 1966 in a fluorescence experiment using a rare earth ion placed in front of a mirror \cite{Drexhage1966}. Over the past decades technological advances have made precise fabrication of nanostructered materials possible, allowing for tailoring the LDOS \cite{Johansen2008, Frimmer2011, Lodahl2004,Lodahl2015}. Engineering the LDOS is attractive since it controls light-matter interaction such as thermal emission, absorption, and spontaneous emission. 

Metamaterials and metasurfaces are nanostructured  three- and two dimensional materials that aim to mimic homogeneous materials and interfaces, but with unconventional material properties. 
Especially metasurfaces and plasmonic lattices have recently attracted interest in the framework of spontaneous emission control due to their guiding properties, broad optical resonances and high field enhancements \cite{Lunnemann2014,Zhen2008, Poddubny2012,Carminati2015}, the possibility of diverging LDOS in hyperbolic metamaterials \cite{Jacob2012}, and the aspect of controlling magnetic and chiral transitions~\cite{Sersic2012,Plum2009}.  
Already soon after the first metamaterials were made, Ruppin and Martin~\cite{Ruppin2004} and K\"astel and Fleischhauer~\cite{Kastel2005} analyzed the classical Drexhage experiment, but envisioning magnetic, and negative index continuous metamaterials to modulate the LDOS. This thought experiment is interesting for a few reasons: First, metamaterials are often proposed with near-field applications in mind, such as super-resolution imaging, or cloaking. Yet, whether a  metamaterial medium built out of, for instance, magnetically polarizable scatterers, acts as a magnetic medium, is usually tested from the far field with just a single input wave vectors. Instead, a more comprehensive test would be to  measure the local density of states, since Drexhage's effect incorporates the different phase upon reflection~\cite{Ruppin2004,Kastel2005}, and sums over all wave vectors. Second, since an emitter is a point-like probe, approaching it to a metamaterial while measuring the lifetime is a direct method to probe at which source-material separations the effective medium approximation holds despite the inherently discrete geometry of metamaterials. These questions have, to the best of our knowledge, not been addressed previously.

To address these issues we present a semi-analytical point-dipole model, that allows for swift calculations of the LDOS of lattices of arbitrary electric, magnetic and bianisotropic  dipolar scatterers. We utilize the method on two types of lattices consisting of isotropically scattering particles with an electric and a magnetic response, respectively. By comparing the Drexhage effect of the two lattices we explore the validity of treating a surface of subdiffractive pitch, and composed of strong electric and magnetic scatterers as a homogenized electric or magnetic mirror. Furthermore, regimes in which the materials may suitably be treated as a homogeneous material are identified. We spectrally resolve the LDOS in regions in the lattice plane, revealing an increased LDOS by coupling to guided lattice modes.

\section*{Results}

\subsection*{Theoretical framework}
The optical response to plane wave excitation of 2D periodic lattices of electric polarizabilities has previously been reviewed  by de Abajo \cite{GarciadeAbajo2007}. An extension to the full magneto-electric case was presented in \cite{Lunnemann2013, Kwadrin2014, Lunnemann2014}. In the following we shall use results derived in \cite{Lunnemann2014} to which we refer the reader for further details. We consider  a 2D periodic lattice of point scatterers in the dipole approximation positioned at $\bm{R}_{mn}=m\bm{d}_1+n\bm{d}_2$, where $m$ and $n$ are integers, and $\bm{d}_{1}$ and $\bm{d}_{2}$ are the real space basis vectors.  
Previous work, based on finite difference time domain simulations\cite{Maier2003a} and quasistatic multipole theory\cite{Park2004}, has shown that the dipole approximation is warranted for $d_i>3b$, where $b$ is the radius of the spheres.
Each particle is described by a polarizability tensor, $\tensor{\alpha}$, that relates the induced electric and magnetic dipole moment, $\bm{\mu}_e$ and $\bm{\mu}_m$, to a driving electric and magnetic field
$\bm{E}$ and $\bm{H}$
according to \cite{Lindell1994,Sersic2011,Lunnemann2014}
\begin{equation}
\begin{pmatrix}
  \bm{p} \\
  \bm{m}  \\
\end{pmatrix}=\alphat \begin{pmatrix}
  \bm{E} \\
  \bm{H}  \\
\end{pmatrix}\label{eq:polMat}.
\end{equation}
 For ease of notation we use a rationalized unit system as described in ref. \cite{Sersic2011} where e.g. $|\bm{E}|/|\bm{H}|=1$ for a plane wave. We note that $\alphat$ is subject to symmetry constraints and must be made electrodynamically consistent, bound by the optical theorem. This is achieved by addition of radiation damping, $\alphat^{-1} = \alphat_{0}^{-1}-2ik^3\mathbb{I}/3$, to the electrostatic polarizability $\alphat_0$ which can for instance be derived from an LC model. Here $.^{-1}$ denotes matrix inversion, $k$ denotes the wave number, $\mathbb{I}$ is the 6-dimensional identity tensor {\cite{Sersic2011}}.
 The magnetoelectric static polarizability is decomposed as
\begin{equation}
\alphat_0=\mathcal{L}(\omega)
\begin{pmatrix}
  {\alphat}_{0}^{(EE)} & \alphat_{0}^{(EH)} \\
  {\alphat}_{0}^{(HE)} & \alphat_{0}^{(HH)}
\end{pmatrix},\label{eq:staticPol}
\end{equation}
where each matrix element is a $3\times3$ dimensionless matrix. The diagonals ${\alphat}_{0}^{(EE)} (\alphat_{0}^{(HH)})$ reflect a purely electric (magnetic) response, whereas   the off-diagonal tensors $\alpha_{0}^{(EH)}$ ($\alpha_{0}^{(HE)}$), describe bianisotropy, such as  the electric response to magnetic fields (and vice versa). $\mathcal{L}(\omega)$ is a Lorenzian prefactor, typical for a plasmon resonance,
\begin{equation}
\mathcal{L}(\omega) = V\frac{\omega_0^2}{\omega_0^2-\omega^2-i\omega\Gamma}
\end{equation}
with resonance frequency $\omega_0$, Ohmic damping $\Gamma$ and amplitude governed by the volume of the scatterer $V$.

The induced dipole moment on a scatterer at the origin $\bm{R}_{00}$ is  set by the sum of the incident field and the field of all other dipoles in the lattice  \cite{Lunnemann2014}
\begin{equation}
\begin{pmatrix}
\bm{p}_{00}\\\bm m_{00}
\end{pmatrix}=\left[\tensor{\alpha}^{-1}-\Gneq(\kparv,0)\right]^{-1}
\begin{pmatrix}
\bm{E}\sub{in}\\
\bm{H}\sub{in}
\end{pmatrix},\label{eq:dipCont}
\end{equation}
where $\kparv$ is the parallel momentum of the incident plane wave,
\begin{equation}
\Gneq(\kparv,\bm{r})=\sum_{m\neq 0,n\neq 0} \tensor{G}^{0}(\bm{R}_{mn}-\bm{r}) e^{i\kparv\cdot\bm{R}_{mn}}\label{eq:latticeSum}
\end{equation}
and  $\tensor{G}^{0}(\bm{R}_{mn}-\bm{r})$  is the $6\times 6$ dyadic Green function of the medium surrounding the lattice. For our case, we shall assume the surrounding medium to be vacuum.

Calculating the LDOS in front of the lattice requires evaluating the scattered field arising from a single point source, instead of from a plane wave of definite parallel momentum.
One approach would be to expand the field of the dipole in its parallel  plane waves $\bm{\kpar}$  \cite{Novotny2008}. However, the  resulting $\kparv$-integral unfortunately converges poorly, especially for small distances to the dipole source \cite{Capolino2007}. 
Instead we shall use a technique referred to as the array scanning method \cite{Capolino2007}.
We consider a single point source dipole with a point current $\bm{j}$ at position $\bm{r}_0$, $\bm{j}(\bm{r})=-i\omega\delta(\bm{r}-\bm{r}_0)\bm{j}$. Here we use the notion of point current in a more generalized magnetoelectric context, where $\bm{j}$ is a 6-element vector describing both the  electric and magnetic dipole, i.e. $\bm{j}=-i\omega(\bm{\mu}_e,\bm{\mu}_m)^\top$. We may synthesize this single point source by summing infinite  phased arrays of point sources:
\begin{equation}
\widetilde{\bm{j}}(\bm{r}')=\bm{j}\sum_{m,n}\delta\left[\bm{r}'-(\bm{r}_0-\bm{R}_{mn})\right]e^{i\kparv\cdot\bm{R}_{mn}}.
\end{equation}
The original single source current is recovered from the phased array as
\begin{equation}
\bm{j}(\bm{r}')=\frac{\mathcal{A}}{(2\pi)^2} \int_{BZ}\widetilde{\bm{j}}(\bm{r}') \mathrm{d}\kparv,\label{eq:pointSource}
\end{equation}
where $BZ$ denotes the Brillouin zone and $\mathcal{A}$ is the real-space unit cell area. We denote all quantities related to the phased array with a tilde. The incident field at the origin, generated by the phased array, is found by propagating the fields from each dipole in the phased array. We get
\begin{align}
\begin{pmatrix}
\widetilde{\bm{E}}\sub{in}(0)\\
\widetilde{\bm{H}}\sub{in}(0)\\
\end{pmatrix}&=
\Gsum(\kparv,-\bm{r}_0)
\begin{pmatrix}
\bm{\mu}_e\\
\bm{\mu}_m\\
\end{pmatrix} \label{eq:Einb}
\end{align}
where we defined $\Gsum(\kparv,\bm{r})=\Gneq(\kparv,\bm{r})+ \tensor{G}^{0}(\bm{R}_{00}-\bm{r}) e^{i\kparv\cdot\bm{R}_{00}}$ that acts as a field propagator of an array of dipole source.

The induced dipole moment of the scatterer at the origin, driven by the phased array,  is found using \cref{eq:dipCont} and \cref{eq:Einb}.
\begin{equation}
\begin{pmatrix}
\widetilde{\bm{p}}_{00}\\
\widetilde{\bm{m}}_{00}\\
\end{pmatrix}=\frac{1}{\tensor{\alpha}^{-1}-\Gneq(\kparv,0)}\Gsum(\kparv,-\bm{r}_0)
\begin{pmatrix}
\bm{\mu}_e\\
\bm{\mu}_m\\
\end{pmatrix}.
\end{equation}
Similar to \cref{eq:Einb}, we may evaluate the scattered field at a position $\bm{r}$ by multiplying the induced dipole with $\Gsum(\kparv,\bm{r})$ giving
\begin{equation}
\begin{pmatrix}
\widetilde{\bm{E}}\sub{scat}(\bm{r})\\
\widetilde{\bm{H}}\sub{scat}(\bm{r})
\end{pmatrix}=\Gsum(\kparv,\bm{r})\begin{pmatrix}
\widetilde{\bm{p}}_{00}\\
\widetilde{\bm{m}}_{00}\\
\end{pmatrix}\label{eq:EscatPhas}.
\end{equation}
The scattered field from the original single dipole source is found by integrating the scattered field, generated by the phased array, over the entire Brillouin zone:

\begin{equation}
\begin{pmatrix}
\bm{E}\sub{scat}(\bm{r})\\
\bm{H}\sub{scat}(\bm{r})
\end{pmatrix}={\tensor{\mathbb{G}}}\sub{tot}(\bm{r}_0,\bm{r})
\begin{pmatrix}
\bm{\mu}_e\\
\bm{\mu}_m
\end{pmatrix}\\
\label{eq:EscatDipole},
\end{equation}
where 
\begin{equation}
{\tensor{\mathbb{G}}}\sub{tot}(\bm{r}_0,\bm{r})\equiv\frac{\mathcal{A}}{(2\pi)^2}\int_{BZ}\Gsum(\kparv,\bm{r})
 \frac{1}{\tensor{\alpha}^{-1}-\Gneq(\kparv,0)}\Gsum(\kparv,-\bm{r}_0)\mathrm{d}\kparv.
\label{eq:integrant}
\end{equation}
Using \cref{eq:EscatDipole}, the decay rate, $\gamma(\bm{r}_0)$, of an emitter relative to the decay rate, $\gamma\sub{vac}(\bm{r}_0)$, in vacuum is calculated as \cite{Novotny2008}:
\begin{equation}
\frac{\gamma(\bm{r}_0)}{\gamma\sub{vac}}=1+\frac{3}{2k^3}\mathrm{Im}\left[
\begin{pmatrix}
\bm{\hat{\mu}_e}\\
\bm{\hat{\mu}_m}
\end{pmatrix}^{\dag}
{\tensor{\mathbb{G}}}\sub{tot}(\bm{r}_0,\bm{r}_0) \begin{pmatrix}
\bm{\hat{\mu}_e}\\
\bm{\hat{\mu}_m}
\end{pmatrix}\right],\label{eq:decayrate}
\end{equation}
where $^\dag$ is the conjugate transpose and $\hat{\bm{\mu}}_e$ ($\hat{\bm{\mu}}_m$) is the normalized electric (magnetic) dipole moment. In this work we will solely consider electric dipole transitions as source ($\hat{\bm{\mu}}_m=0$). 
We note that the computation of the summation in \cref{eq:latticeSum} is carried out using Ewald summation \cite{Linton2010} described in Supplementary material, and details of the integral in \cref{eq:integrant} is computed in practice are described in Methods.
Moreover, while we only consider a single magneto-electric dipole mode of the scatterers, the model may easily be extended to treat stacked lattices as well as complex unit cells consisting of different scatterers,  to mimic multipolar resonances\cite{Kwadrin2014}. Also, more advanced methods for retrieving the polarizability, e.g. surface integral equations\cite{Arango2013}, may be used as input.

\subsection*{Numerical examples}
As examples we shall consider non-diffractive square lattices of strong scatterers.
\begin{table}
\centering
\begin{tabular}{crl}
\hline
\textbf{Parameter} & \textbf{Value} & \textbf{Description}\\\hline
{$d_1=d_2\equiv d$} & $300\unit{nm}$ & Lattice constant.\\
$\lambda_0=2\pi c/\omega_0$ & $1.5\unit{\micro\meter}$ & Res. wavelength of  particles.\\
$\Gamma$ & $83\unit{THz}$ & Ohmic damping in particles\cite{Lunnemann2013}.\\
$V$ & $(90\unit{nm})^3$ & Volume of scatterer.\\\hline
\end{tabular}
\caption{\label{tab:params} Used parameters for the calculations}
\end{table}
We calculate the LDOS near lattices of two types of scatterers: (1) Scatterers with an isotropic electric response (i.e. plasmonic spheres) by setting
$\tensor{\alpha}_{0}^{(EE)}=\mathcal{L}(\omega)
\mathbb{I}$,
and ${\tensor{\alpha}}_{0}^{(EH)}={\tensor{\alpha}}_{0}^{(HE)}={\tensor{\alpha}}_{0}^{(HH)}=\tensor{0}$, and (2) Scatterers with an isotropic magnetic response by setting $\tensor{\alpha}_{0}^{(HH)}=\mathcal{L}(\omega)
\mathbb{I}$,
and $\tensor{\alpha}_{0}^{(EH)}=\tensor{\alpha}_{0}^{(HE)}=\tensor{\alpha}_{0}^{(EE)}=\tensor{0}$. All  parameter values used are presented in Table \ref{tab:params}. The parameters are chosen so that the electric scatterers match the polarizability, extinction cross section and albedo  found experimentally  for plasmonic scatterers at telecom frequencies (extinction cross section $0.38\unit{\micro\meter}^2$), as studied in depth by Husnik et al.\cite{Husnik2008,Husnik2012}. 

\begin{figure}
\begin{center}
\newlength\figureheight 
\newlength\figurewidth 
\includegraphics[width=.6\columnwidth]{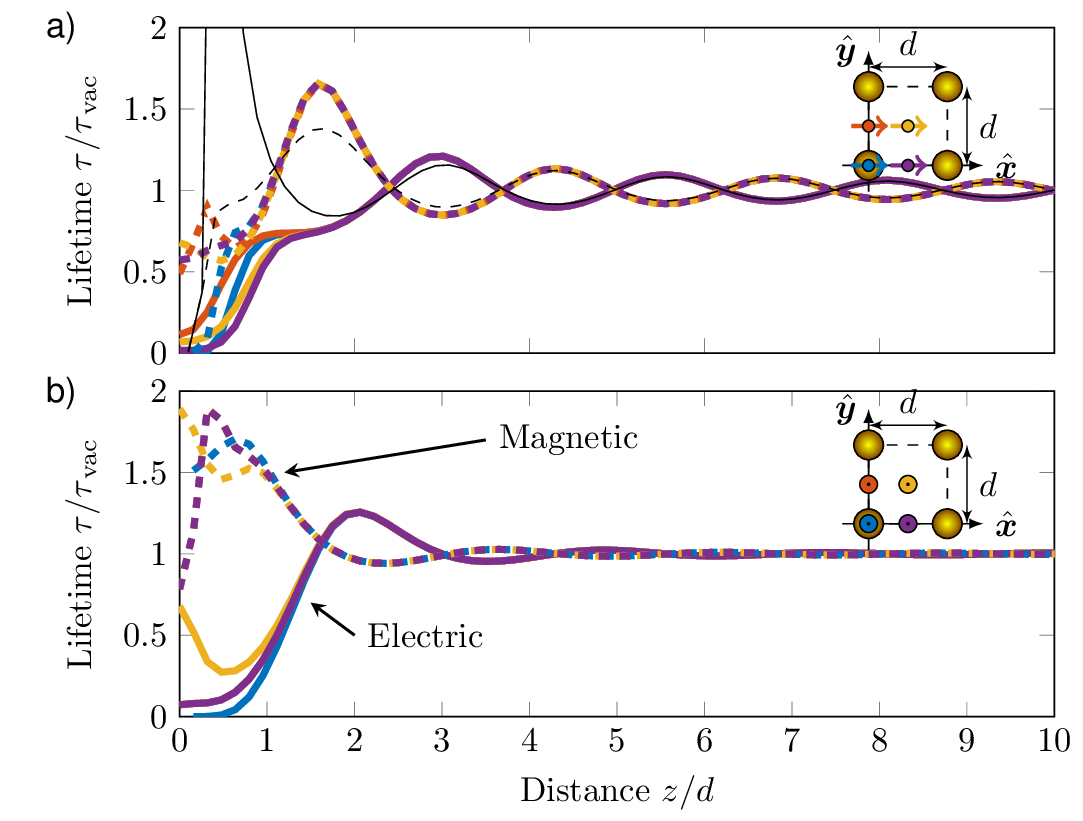}
\caption{Calculated relative lifetime of an electric dipole as a function of distance for the case of a lattice with isotropic magnetic scatterers (dashed) and electric scatterers (solid), at the resonance frequency $\omega_0$ for the four in-plane positions $(x,y) = (d/2,0)$ (blue), $(x,y) = (0,d/2)$ (red), $(x,y) = (d/2,d/2)$ (yellow), and $(x,y) = (d/2,0)$ (purple) as depicted in the inset. a) Dipoles oriented along $\unitvec{x}$ parallel to the lattice. b) Dipoles oriented perpendicular to the surface. Thin black lines are calculated lifetimes assuming a homogeneous planar interface.\label{fig:fig1}} 
\end{center}
\end{figure}

The calculated LDOS modulation (plotted as predicted fluorescence lifetime normalized to lifetime in vacuum) as a function of distance is presented in \cref{fig:fig1}, for an electric dipole source positioned at the four symmetry points $\bm{r}_{||}=(0,0)d/2$, $\bm{r}_{||}=(1,0)d/2$, $\bm{r}_{||}=(1,1)d/2$, and $\bm{r}_{||}=(1,1)d/2$, oriented parallel to the lattice plane along $\unitvec{x}$, \cref{fig:fig1}a), and perpendicular to the lattice plane along $\unitvec{z}$, \cref{fig:fig1}b).
The relative lifetimes oscillate as a function of distance with a periodicity of about $\lambda/2$, as encountered in typical Drexhage-type experiments\cite{Drexhage1966,Johansen2008,Amos1997,Lunnemann2013a}. 
Comparing the electric versus magnetic lattices, we note that the oscillations in lifetime are $\pi$ out of phase. A similar effect was predicted  by Ruppin and Martin~\cite{Ruppin2004} for hypothetical `magnetic mirrors', i.e.,  for reflection at a medium  that presents $\mu=-\infty, \epsilon=1$,  as opposed to $\epsilon=-\infty,\mu=1$ for a normal electric mirror. In their work, the difference is  associated with a $\pi$ difference in Fresnel reflection coefficients that appears  when interchanging  magnetic permeability and electric permittivity. The calculated Drexhage oscillations, and their reversal in phase with exchanging the nature of the scatterers hence confirms that electric (magnetic) particle lattices act as effective electric (magnetic) reflective interfaces.

Considering the case of an electrically scattering lattice, solid lines in \cref{fig:fig1}, we note, that for distances beyond $\sim 2d$, or equivalently about $\lambda/3$, the lifetimes at the four different positions are indiscernible. Above this distance, the lattice is well approximated as an effective homogeneous material, as often assumed~\cite{Ruppin2004,Kastel2005,Jacob2012}. To qualify this statement further,  we calculated the angle-dependent \textit{far field}  reflection coefficients (using  \cref{eq:dipCont} and equation (15) in ref. \cite{Lunnemann2013}). These reflection constants can be used as input to textbook expressions for the LDOS near a homogeneous interface \cite{Amos1997,Novotny2008}, which for electric sources perpendicular, respectively parallel to an interface read
\begin{equation*}
\rho_{\perp,E}=\frac{3}{2}\mathrm{Im}\int_0^\infty [1-r_p (k_{||})e^{-2ik_zd}] \frac{ik_{||}^3}{k_z}dk_{||}
\end{equation*}
and
\begin{equation*}
\rho_{||,E} =  \frac{3}{4}\mathrm{Im}\int_0^\infty\Big\{ [1+r_s (k_{||})e^{-2ik_zd}] + (1-k_{||}^2)[1+r_p (k_{||})e^{-2ik_zd} ]\Big\}
\frac{ik_{||}}{k_z}dk_{||}. 
\end{equation*}
Here $k_z(k^2-\kpar^2)^{1/2}$ and  $r_{s,p}$  represent the $s$- and $p$-reflection coefficient, and  integrating up to  $k_{||}=k$ accounts for all far-field reflection effects. We find excellent agreement for distances beyond a few lattice constants. This delineates the validity of using far field measurements to obtain  effective material parameters.  Furthermore, the notion  of "effective material parameter" should be read as meaning that the medium is fully quantified by its far-field reflection for all angles, irrespective of the question if these reflection constants are consistent with any $\epsilon$ and $\mu$.

For closer distances than $\sim 2d$, the discrete nature of the lattice is revealed in the position dependence of the decay. For all four positions, the lifetime rapidly decreases for short decreasing distances. Naturally, very close to a scattering sphere we expect a decrease associated with the near field of a single sphere. This should occur for ranges of order 50 nm ($a/5$)~\cite{Mertens2007}. 

\begin{figure}
\begin{center}
\includegraphics[width=.8\columnwidth]{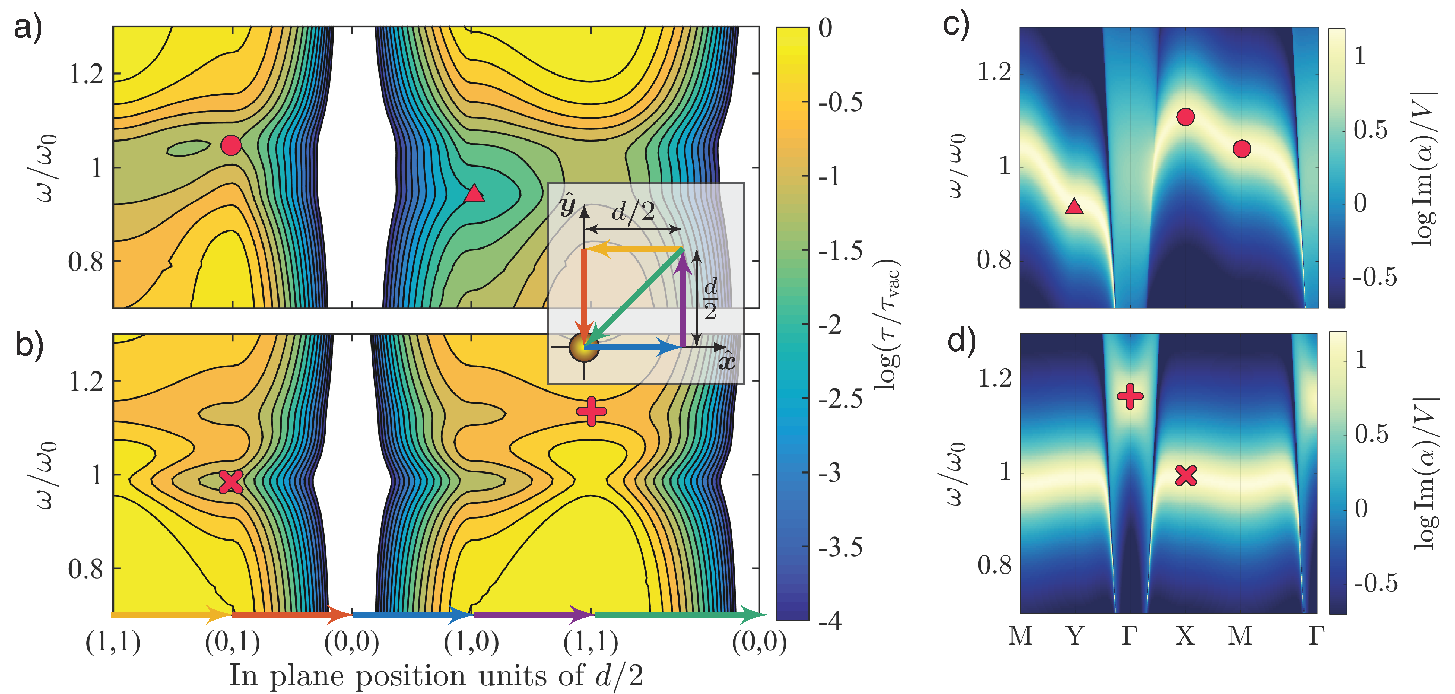}
\caption{Spectral dependence of the relative lifetime of a dipole emitter placed in the plane ($z=0$) of an electric isotropically scattering lattice, as a function of emission frequency and in-plane position along straight paths as illustrated by the colored arrows and the inset. a) Dipole along $\unitvec{x}$ parallel to the lattice. b) Dipole along $\unitvec{z}$ perpendicular to the lattice. Markers illustrate different positions of the emitter. The inferred coupling to guided modes, associated with these positions, are marked on the calculated dispersion of the two modes with c) induced dipoles parallelle to the plane of the lattice along $\unitvec{x}$ and d) induced dipoles perpendicular to the lattice plane.\label{fig:LDOSvsFreq}}
\end{center}
\end{figure}

For intermediate distances we identify a third effect, namely coupling to guided modes in the lattice \cite{Lunnemann2014}. To  investigate contributions from guided modes we calculated the relative lifetime as a function of its emission frequency and in-plane position for a parallel and perpendicular dipole positioned in the plane of an electric isotropically scattering lattice, presented in \cref{fig:LDOSvsFreq}c)-d).
Firstly, we note, that for a perpendicular (parallel) dipole positioned in the plane of the lattice, all electric field components in the plane are perpendicular (parallel) to the lattice plane. Hence we expect coupling to modes with induced dipoles being purely perpendicular (parallel) to the plane.
Considering the case of a parallel dipole (\cref{fig:LDOSvsFreq}d)) we firstly notice that close to the scattering element at $(x,y)=(0,0)$, the lifetime drastically decreases owing to the $1/r^3$ scaling of the near field  of  the scatterer. Elsewhere, distinct bands of reduced lifetimes are resolved for frequencies different from the resonance frequency of the individual scatterer. This indicates that the source dipole couples not simply to the individual scattering elements, rather it couples to a guided lattice mode that is frequency dispersive. E.g. near $(x,y)=(0,d/2)$, marked with a red circle, a significant reduction of the lifetime occurs for blue shifted frequencies relative to the single particle resonance frequency ($\omega_0$).  Symmetry of the lattice and the field lines of a dipole imply that the band arises from coupling to a longitudinal in-plane mode (LI) where the induced dipoles are arranged in a head to head configuration along $\unitvec{x}$. 
This is confirmed from the calculated dispersion of the lattice mode with induced dipoles parallel to $\unitvec{x}$, presented in \cref{fig:LDOSvsFreq}a) (for details on the calculation of the modal dispersion we refer to Ref. \cite{Lunnemann2014}). At points $\mathrm{X}\equiv \kparv = (\pi/d,0)$ and $\mathrm{M}\equiv \kparv = (\pi/d,\pi/d)$, the mode is blueshifted with a flat slope thus giving rise to a large LDOS.
Similarly, near $(x,y)=(d/2,0)$, marked with a triangle, a reduction is seen to occur for red shifted frequencies corresponding to a transverse in-plane mode with $\kparv=\mathrm{Y}\equiv(0,\pi/2)$.
In the case of a dipole perpendicular to the lattice,  (\cref{fig:LDOSvsFreq}d)), the calculated lifetime is symmetric about $(0,0)$ owing to the four-fold rotational symmetry of the lattice. Two bands appear near $(0,d/2)$ and $(d/2,0)$ with one being slightly red shifted, the other blue shifted relative to the resonance frequency $\omega_0$.  
Comparing with the calculated mode with induced dipole momements perpendicular to the lattice plane, shown in \cref{fig:LDOSvsFreq}b), we conclude that the red shifted band is associated with coupling to a transverse guided mode with the induced dipoles perpendicular to the lattice, 
while the blue shifted resonance arise from coupling to a non-guided mode with wavevectors near above the light line. Coupling to this leaky mode is only achieved close to the lattice, since only in the near field of a radiating dipole does it contain wave vectors parallel to its dipole moment. Due to symmetry at $(1,1)d/2$, marked by $+$, only coupling to the blue shifted non-guided mode with vanishing in-plane wavevectors near the point $\kparv=\mathrm{\Gamma}\equiv(0,0)$, remains.

\section*{Discussion and conclusion}
In conclusion, we have presented a simple point dipole method using the array scanning method for calculating the LDOS of an arbitrary magnetoelectric infinite 2D lattice.  
The primary motivation to tackle this problem was to assess in how far analyzing a metamaterial as effectively homogeneous is reasonable in an actual scenario where it interacts with a localized object in its near field. As example, we calculated the lifetime of a dipole in front of electric and magnetic isotropically scattering spheres. 
We found that a lattice of magnetic scatterers shows characteristic oscillations of the LDOS as a function of distance, shifted in phase compared to those at an electric scattering lattice.   This confirms that a metamaterial can appear as a magnetic mirror also in "Drexhage" experiments that are not limited to probing by a single far field incidence angle,  as was first proposed by Ruppin and Martin~\cite{Ruppin2004} and K\"astel and Fleischhauer~\cite{Kastel2005}. Our results reveal that for distances beyond $2d\sim\lambda/3$, the surfaces can be well approximated as an effective homogenous interface, with electric and magnetic properties taken from far field reflection constants.
For somewhat shorter distances the lifetime shows a dependence on both in-plane position and frequency that is due to the discrete nature of the lattice, and coupling to lattice guided modes, which is not captured by far field reflection constants. At even shorter distances comparable to feature sizes of the scatterer, where microscopic detail matters, equation (7) of our work remains valid, however, the dipole approximation breaks down. Microscopically, one could use a full-wave solver (FDTD, COMSOL) for every wave vector in the integral in \cref{eq:pointSource}. In practice, however, this leads to an impractical computational burden. As an intermediate, and more tractable, approach we propose to improve the microscopic detail captured by our model by using multiple dipoles to describe a single scatterer, instead of using single dipoles \cite{Kwadrin2014}.

These results are of fundamental interest to the question how one probes the range of validity of effective medium parameters in near field geometries. 
Furthermore, our method is excellently suited for emitters with an excited state subject to competing radiative decay pathways with electric, magnetic, and mixed character \cite{Taminiau2012, Karaveli2013}, where the calculated LDOS for the magnetic and electric transitions may be used as coefficients in the rate equations for the density of states of the emitter. Finally, our method can be easily extended to diffractive plasmonic systems, arbitrarily complex unit cells~\cite{Kwadrin2014}, multilayered unit cells, and  bi-anisotropic or hyperbolic metasurfaces.

\section*{Methods}\label{sec:methods}
The integrant in \cref{eq:integrant} typically contains sharp features over the $\kparv$-plane, so an adaptive numerical integration is crucial. Furthermore, since 
\begin{equation}
k_z=\sqrt{k^2-|\kparv|^2}
\end{equation}
there is a branchpoint at $|\kparv|=k$.  Since the routine method of performing  the integration over $\kparv$ into the complex plane \cite{Paulus2000} would require complex $\kparv$-lattice sums, we avoid it.  Instead we split the integration into two different domains. For  $\kpar$ within the light cone we use polar coordinates
\begin{equation}
\kparv\in\{(\phi,\kpar)\in\mathbb{R}| 0\leq\phi\leq\pi \quad\wedge\quad \kpar<k\}.
\end{equation}
Outside the light cone cartesian coordinates $(k_x,k_y)$ are used for  
\begin{equation}
%\begin{split}
\{(k_x,k_y)\in\mathbb{R}| 0\leq k_x\leq\frac{\pi}{d_1} \wedge %\\
 \mathrm{Re}\left(\sqrt{k^2-k_x^2}\right) \leq k_y\leq\frac{\pi}{d_2}\}
%\end{split}
\end{equation}
were used. Rather than computing  all 36 tensor elements we directly calculated
\begin{equation}
\int_{BZ}\frac{3}{2}k^{-3}\mathrm{Im}\left[
\begin{pmatrix}
\bm{\hat{\mu}}_e\\
\bm{\hat{\mu}}_m
\end{pmatrix}^{\dag}
\Gsum\sub{tot}(\kparv,\bm{r},\bm{r}_0)
\begin{pmatrix}
\bm{\hat{\mu}}_e\\
\bm{\hat{\mu}}_m
\end{pmatrix}\right]
\mathrm{d}\kparv,\label{eq:numIntegrant}
\end{equation}
where 
\begin{equation}
\Gsum\sub{tot}(\kparv,\bm{r},\bm{r}_0)\equiv\Gsum(\kparv,\bm{r})\frac{1}{\tensor{\alpha}^{-1}-\Gneq(\kparv,0)}\Gsum(\kparv,-\bm{r}_0).
\end{equation}

%\bibliography{/Users/kaqmak/Dropbox/articles/bibtex/Papers-LatticeLDOS.bib}

\section*{Acknowledgments}
This work is part of the research program of the Stichting
voor Fundamenteel Onderzoek der Materie (FOM), which is
financially supported by the Nederlandse Organisatie voor
Wetenschappelijk Onderzoek (NWO). PL acknowledges the Carlsberg Foundation and the Danish Research Council for Independent Research (Grant No. FTP 11-116740).

\section*{Author contributions}
P.L. and A.F.K. were responsible for the original research concept and physical interpretation. A.F.K was mainly responsible for the theoretical derivations while P.L. wrote the computer code  manuscript with the help of Q.Z. All authors reviewed the manuscript.

\section*{Additional information}
The authors declare no competing financial interests.

\newpage
\renewcommand{\thesection}{S\arabic{subsection}}%... from subsections
\renewcommand{\theequation}{S\arabic{equation}}%... 
\setcounter{equation}{0}
\section*{Supplementary material}
\subsection*{Sums of magneto-electric Dyadic Greens function}\label{sec:appendix1}
The sum presented in equation (5), requires special attention since it converges poorly. The problem has been treated extensively in ref. 1 and utilizes a technique pioneered by P. Ewald. The technique consists in splitting a poorly convergent sum into two convergent terms, $\Gsum^{(1)}$ and $\Gsum^{(2)}$, which are exponentially convergent. Specifically, considering the sum
\begin{equation}
\Gamma(\bm{k}_{||},\bm{r})=\sum_{m,n}
G^0({\bm{R}}_{mn}-\bm{r})e^{i\bm{k}_{||}
\cdot \bm{R}_{mn}}\label{eq:scalLatticeSum}
\end{equation}
where the scalar Green function is
\begin{equation}
G^0({\bm{R}}_{mn}-\bm{r})=\frac{e^{ik
|\bm{R}_{mn}-\bm{r}|}}{|\bm{R}_{mn}-\bm{r}|}.
\end{equation}
we may rewrite this as
\begin{equation} \sum_{m,n}
\frac{e^{ik
|\bm{R}_{mn}-\bm{r}|}}{|\bm{R}_{mn}-\bm{r}|}
e^{i\bm{k}_{||}\cdot\bm{R}_{mn}}=\Gamma^{(1)}+\Gamma^{(2)}.
\end{equation}
Here
\begin{subequations}
\begin{equation}
\Gamma^{(1)}=\frac{\pi}{{\cal{A}}}\sum_{\tilde{m}\tilde{n}}\left\{
\frac{e^{i(\bm{k}_{||}+g_{\tilde{m}\tilde{n}})\cdot
\bm{r}_{||}}}{k^z_{\tilde{m}\tilde{n}}}\right.
\cdot\left[
e^{ik^z_{\tilde{m}\tilde{n}}|z|}\mathrm{erfc}\left(\frac{k^z_{\tilde{m}\tilde{n}}}{2\eta}+|z|\eta\right)\right.
+
\left.\left. e^{-ik^z_{\tilde{m}\tilde{n}}|z|}\mathrm{erfc}\left(\frac{k^z_{\tilde{m}\tilde{n}}}{2\eta}-|z|\eta\right) \right]\right\}
\label{eq:cylindricalsum}
\end{equation}
and 
\begin{equation}
\Gamma^{(2)}=%\frac{1}{2}
\sum_{mn}\left\{\frac{e^{i\bm{k}_{||}\cdot\bm{R}_{mn}}}{2\rho_{mn}} 
\cdot\left[ e^{ik\rho_{mn}}\mathrm{erfc}\left(\rho_{mn}\eta
+\frac{ik}{2\eta}\right) \right.\right.
+ \left.\left. e^{-ik\rho_{mn}}\mathrm{erfc}\left(\rho_{mn}\eta
-\frac{ik}{2\eta} \right) \right]\right\},
\label{eq:sphericalsum}
\end{equation} 
\end{subequations}
where we used $\bm{r}=(\bm{r}_{||},z)$,
$k=\omega/c$, $k^z_{\tilde{m}\tilde{n}}=
\sqrt{k^2-|\bm{k}_{||}+\tensor{g}_{\tilde{m}\tilde{n}}|^2}$, and
$\rho_{mn}=|\bm{R}_{mn}-\bm{r}_{||}|$. Convergence of \cref{eq:sphericalsum} and \cref{eq:cylindricalsum} follows from the asymptotic expansion of the  error function revealing $z\,\mathrm{erfc}(z)\sim \exp(-z^{2})$ for $z\rightarrow\infty$.\cite{Linton2010}
The parameter $\eta$ can be chosen for optimal convergence, and
should be set around $\eta=\sqrt{\pi}/{a}$, where $a$ is the lattice
constant. Naturally, the cut off  for the summation over $m$ and $n$ must be chosen at least bigger than the number of propagating grating diffraction orders one expects.For our calculations on metamaterials, with essentially no grating orders, i.e., $ka\leq 2\pi$, we already obtained converged lattice sums for  $|m,n|\leq 5$.  

The dyadic lattice sums in equation (5) are easily generated by noting that the
scalar Green function 
\begin{equation}
G(\bm{r},\bm{r'})=\frac{\exp\left({ik|\bm{r}-\bm{r}'|}\right)}{|\bm{r}-\bm{r}'|}
\end{equation}
 sets the dyadic
Green function via
\begin{equation}
\tensor{G}^0(\bm{r}-\bm{r}') =
\begin{pmatrix}
\mathbb{I} k^2 +\nabla\otimes\nabla &
-ik\nabla \times  \\
 ik\nabla \times   &  \mathbb{I} k^2
+\nabla\otimes\nabla \\
\end{pmatrix} G(\bm{r},\bm{r}')
\label{eq:derive}
\end{equation}
where $\mathbb{I}$ indicates the $3\times 3$ identity matrix and $\otimes$ denotes the outer product.
The derivatives can be simply pulled into each exponentially
convergent sum to be applied to each term separately, and are most
easily implemented in practice by noting that the sum
$\Gamma^{(2)}$ only depends on radius in spherical coordinates
$\rho_{mn}$, while the sum in $\Gamma^{(1)}$ only depends on radius and height in cylindrical coordinates. For these coordinate systems the differential operator in \cref{eq:derive} take particularly simple forms. For spherical coordinates this form reads
\begin{subequations}
\begin{equation} (\mathbb{I} k^2 +\nabla\nabla )F(r) =  
\mathbb{I}\left[k^2  F(r) + \frac{1}{r}\frac{d}{dr} F(r)\right]
+
\begin{pmatrix}
x^2 & xy & xz \\
xy & y^2 & yz \\
xz & yz  & z^2 \\
 \end{pmatrix}
 \frac{1}{r}\frac{d}{dr}\left[ \frac{1}{r}\frac{d}{dr} F(r)\right]
 \end{equation}
 and
 \begin{equation}
 -ik\nabla \times F(r)  =   ik \begin{pmatrix}
0 & z & -y    \\
-z & 0 & x \\
y & -x  & 0 \\
 \end{pmatrix}\frac{1}{r}\frac{d}{dr} F(r),
\end{equation}
\end{subequations}
which can be directly applied to the summands in
\cref{eq:sphericalsum}.
 For cylindrical coordinates the differential  form reads
 \begin{subequations}
\begin{multline} 
(\mathbb{I} k^2 +\nabla\otimes\nabla )
e^{i\bm{k}\cdot\rho} g(z)=
\begin{pmatrix}
k^2-k_x^2 & -k_xk_y & 0 \\
 -k_xk_y & k^2-k_y^2 & 0 \\
0 &0  & k^2 \\
 \end{pmatrix}e^{i\bm{k}\cdot\bm{r}_{||}} g(z)\\
+\begin{pmatrix}
0 & 0 & ik_x  \\
 0 & 0 & ik_y \\
ik_x &ik_y  & 0  \\
 \end{pmatrix}e^{i\bm{k}\cdot\bm{r}_{||}} \frac{\mathrm{d} g(z)}{\mathrm{d}z}
+\begin{pmatrix}
0 & 0 & 0  \\
 0 & 0 & 0 \\
0 & 0  & 1  \\
\end{pmatrix}e^{i\bm{k}\cdot\bm{r}_{||}} \frac{\mathrm{d}^2 g(z)}{\mathrm{d}z^2}
\end{multline}
and
\begin{equation}
  -ik\nabla \times e^{i\bm{k}\cdot\bm{r}_{||}} g(z)  =
 \begin{pmatrix}
0 & 0 & -k k_y  \\
 0 & 0 &  kk_x \\
kk_y &-kk_x  & 0  \\
\end{pmatrix}
e^{i\bm{k}\cdot\bm{r}_{||}} g(z) 
+
\begin{pmatrix}
0 & ik & 0  \\
 -ik & 0 & 0 \\
0 & 0  & 0  \\
\end{pmatrix}e^{i\bm{k}\bm{r}_{||}} \frac{d g(z)}{dz}
\label{eq:cyldiff}\end{equation}
 \end{subequations}
which can be directly applied to evaluate the dyadic equivalent of
\cref{eq:cylindricalsum}.

\end{document}